\documentclass[hidelinks,twoside]{tufte-handout}

\usepackage[utf8]{inputenc} 

\usepackage{geometry} 
\usepackage{graphicx} 
\usepackage{caption}
\usepackage[caption=false]{subfig}
\usepackage[export]{adjustbox}

\usepackage{booktabs} 
\usepackage{multirow} 

\usepackage{paralist} 
\usepackage{verbatim} 

\usepackage{array} 
\usepackage{amsmath} 
\usepackage{amsthm} 
\usepackage{amssymb} 
\usepackage{mathtools}

\usepackage{algorithm}
\usepackage[noend]{algpseudocode}

\usepackage{natbib}

\theoremstyle{definition}
\newtheorem{definition}{Definition}

\newtheorem{theorem}{Theorem}
\newtheorem{lemma}{Lemma}
\newtheorem{example}{Example}

\def\LLP/{\emph{LLP supervisor}}

\titleformat{\section}[block]
  {\fontsize{12}{20}\bfseries\sffamily\filcenter}
  {\thesection}
  {1em}
  {\MakeUppercase}
\titleformat{\subsection}[hang]
  {\fontsize{11}{12}\bfseries\sffamily}
  {\thesubsection}
  {1em}
  {}

\makeatletter
\newcommand{\@tufte@print@margin@citation}[1]{%
    \citealp{#1}
}

\renewcommand{\@tufte@normal@cite}[2][0pt]{%
  \let\@temp@last@bibkey\@empty%
  \@for\@temp@bibkey:=#2\do{\let\@temp@last@bibkey\@temp@bibkey}%
  \sidenote[][#1]{%
    \normalsize\normalfont\@tufte@citation@font%
    \setcounter{@tufte@num@bibkeys}{0}%
    \@for\@temp@bibkeyx:=#2\do{%
      \ifthenelse{\equal{\@temp@last@bibkey}{\@temp@bibkeyx}}{%
        \ifthenelse{\equal{\value{@tufte@num@bibkeys}}{0}}{}{and\ }%
        \@tufte@trim@spaces\@temp@bibkeyx
        \@tufte@print@margin@citation{\@temp@bibkeyx}%
      }{%
        \@tufte@trim@spaces\@temp@bibkeyx
        \@tufte@print@margin@citation{\@temp@bibkeyx};\space
      }%
      \stepcounter{@tufte@num@bibkeys}%
    }%
  }%
}

\newcommand{\resetcitations}{%
  \gdef\@tufte@old@bibkeys{}%
}
\makeatother

\title[LLP for the Control of DES: A Tutorial]{Limited Lookahead Policies for the Control of\\Discrete-Event Systems: A Tutorial}
\author[R.H. Moulton, A.J. Marasco, and Karen Rudie]{Richard~Hugh~Moulton,\thanks{\texttt{richard.moulton@queensu.ca}\\\noindent Department of Electrical and Computer Engineering, Queen's University, Kingston ON, Canada} Anthony~J.~Marasco,\thanks{\texttt{anthony.marasco@rmc-cmr.ca}\\\noindent School of Computing, Queen's University, Kingston ON, Canada\\\noindent Department of Electrical and Computer Engineering, Royal Military College of Canada, Kingston ON, Canada}\\and Karen~Rudie\thanks{\texttt{karen.rudie@queensu.ca}\\\noindent Department of Electrical and Computer Engineering and Ingenuity Labs Research Institute, Queen's University, Kingston ON, Canada}}

\begin{document}
\maketitle

\vskip\bigskipamount 
\leaders\vrule width \textwidth\vskip0.4pt 
\vskip\medskipamount 
\nointerlineskip

\begin{abstract}
Some problems in discrete-event systems (DES) model large, time-varying state spaces with complex legal languages. To address these problems, Chung et al.\ introduced limited lookahead policies (LLP) to provide online supervisory control for discrete-event systems. This seminal paper, along with an addendum of technical results, provided the field with a series of very important and powerful results, but in a notationally- and conceptually-dense manner. In this tutorial, we present Chung et al.'s problem formulation for online control and unravel the formal definitions and proofs from their original work with the aim of making the ideas behind limited lookahead accessible to all DES researchers. Finally, we introduce the Air Traffic Control problem as an example of an online control problem and demonstrate the synthesis of LLP supervisors.
\end{abstract}

\vskip\medskipamount 
\leaders\vrule width \textwidth\vskip0.4pt 
\vskip\bigskipamount 
\nointerlineskip

 \setcounter{footnote}{0} 

\section{Introduction}
\newthought{The main problem} in control theory is the problem of providing a control signal to a process to ensure that only desirable behaviour is produced. In discrete event systems (DES) this is done through supervisory control: enabling or disabling events that cause the process to transition from one state to another. Although DES theory provides supervisors that are correct by construction, there are real-world applications whose characteristics make constructing such supervisors infeasible. These applications include processes with a very large state space, processes that are time-varying, and processes whose desirable behaviours are difficult to specify.

Chung, Lin and Lafortune's 1992 paper ``Limited Lookahead Policies in Supervisory Control of Discrete Event Systems'' addressed these challenges by providing a theoretical base for the development of DES supervisors that performed control in an online manner rather than computing a monolithic control policy offline.\cite{Chung1992} This work inspired a number of follow-on works and was recognized with the George S. Axelby Outstanding Paper Award for the \emph{IEEE Transactions on Automatic Control}. Nonetheless, we believe that wider use of these limited lookahead policy (LLP) supervisors has been prevented by the conceptually dense material that must be mastered.

In this tutorial we present Chung et al.'s problem formulation, supervisor characterization and formal proofs of correctness in a way that emphasizes the underlying ideas. Our aim is to make the theoretical results in ``Limited Lookahead Policies in Supervisory Control of Discrete Event Systems'' usable for researchers and practitioners alike. We begin by introducing the \emph{Air Traffic Control (ATC) problem} as an illustrative example of the online control problem we use throughout the tutorial (Section~\ref{sec:ATCproblem}). Next we present Chung et al.'s seminal results (Sections~\ref{sec:DESoperations} through~\ref{sec:CompareSupervisors}) and demonstrate how these results are used to synthesize LLP supervisors (Section~\ref{sec:SynthesizeLLPSupervisor}). Finally, we provide an overview of works that directly extend or apply ideas from the original LLP supervisor (Sections~\ref{sec:LLPextensions} and~\ref{sec:LLPapplications}).

\section{The Air Traffic Control Problem} \label{sec:ATCproblem}
\newthought{We will use the} ATC problem as a concrete example throughout this paper to illustrate the theoretical concepts we are discussing. This problem requires the formulation of a supervisor to control air traffic in the vicinity of an airfield. It is useful because it is conceptually easy to understand and because some of its characteristics make it challenging for traditional DES approaches. 
\begin{figure}[htb]
\centering
\includegraphics[width=\textwidth]{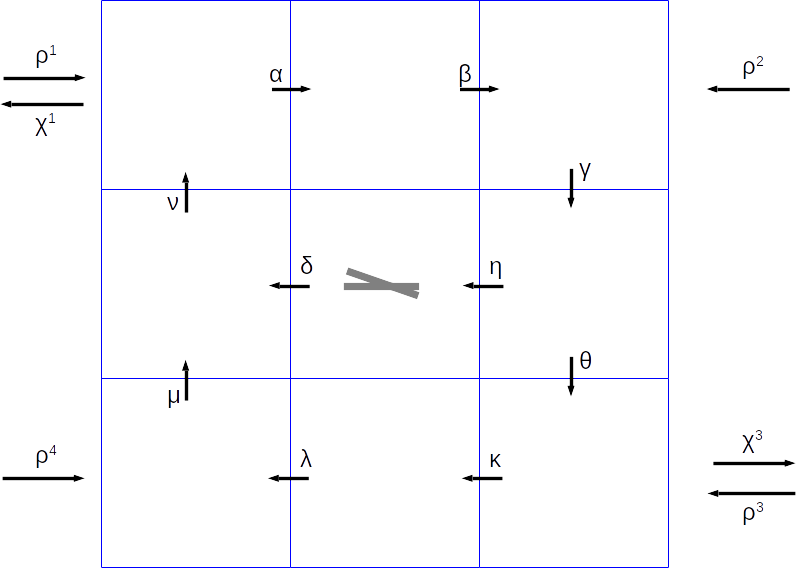}
\caption{The Air Traffic Control problem. The airspace to be controlled is shown: the airfield is in the centre square; aircraft can arrive in any of the four corners, but can only leave from the top left and bottom right corners; and once an aircraft is in the airspace it must travel in a clockwise manner around the airfield. The supervisor that we synthesize must ensure that aircraft can take off and land from this airfield with minimal restrictions while guaranteeing flight safety.}
\label{fig:atcproblem}
\end{figure}

\subsection{The Alphabet}
As a discrete event system model, the ATC problem uses the alphabet $\Sigma = \Sigma_{c}\enspace \dot{\bigcup}\enspace \Sigma_{uc}$. The controllable events, denoted by $\Sigma_{c}$ \eqref{eq:eventsC}, represent aircraft movements into a new section of airspace, as well as take-off ($\delta_i$), landing ($\eta_i$), entering the airfield's vicinity ($\rho^j_i$), and leaving the airfield's vicinity ($\chi^j_i$). The uncontrollable events, denoted by $\Sigma_{uc}$ \eqref{eq:eventsUC}, represent notifications of a pending arrival or take off and are distinguished by a tilde, \textasciitilde.  All events are subscripted with the index of the plane performing the action.
\begin{align}
\Sigma_c = \{&\epsilon, \alpha_i, \beta_i, \gamma_i,\theta_i, \kappa_i, \lambda_i, \mu_i, \nu_i, \eta_i,\nonumber\\
&\delta_i, \rho^1_i, \rho^2_i, \rho^3_i, \rho^4_i, \chi^1_i, \chi^3_i\}\ i \in \mathbb{N}^+ \label{eq:eventsC}\\
\Sigma_{uc} = \{&\tilde{\rho}^{1}_i, \tilde{\rho}^{2}_i, \tilde{\rho}^{3}_i, \tilde{\rho}^{4}_i, \tilde{\delta}_i\}\ i \in \mathbb{N}^+ \label{eq:eventsUC}
\end{align}

In order to simplify the problem, we assume that all events are observable, although subsequent work allowed for unobserved events in the plant (see Section~\ref{sec:LLPextensions}).

\subsection{The Plant}
The plant representing the uncontrolled airspace can be modelled as the synchronous product of multiple subplants. This will allow us to reason about each component of the plant separately, without having to consider the effects of the whole plant.

\begin{figure}[htb]
\centering
\captionsetup{captionskip=0pt,farskip=0pt,nearskip=0pt}

\vspace*{-\abovecaptionskip}

\includegraphics[width=0.8\textwidth,valign=T]{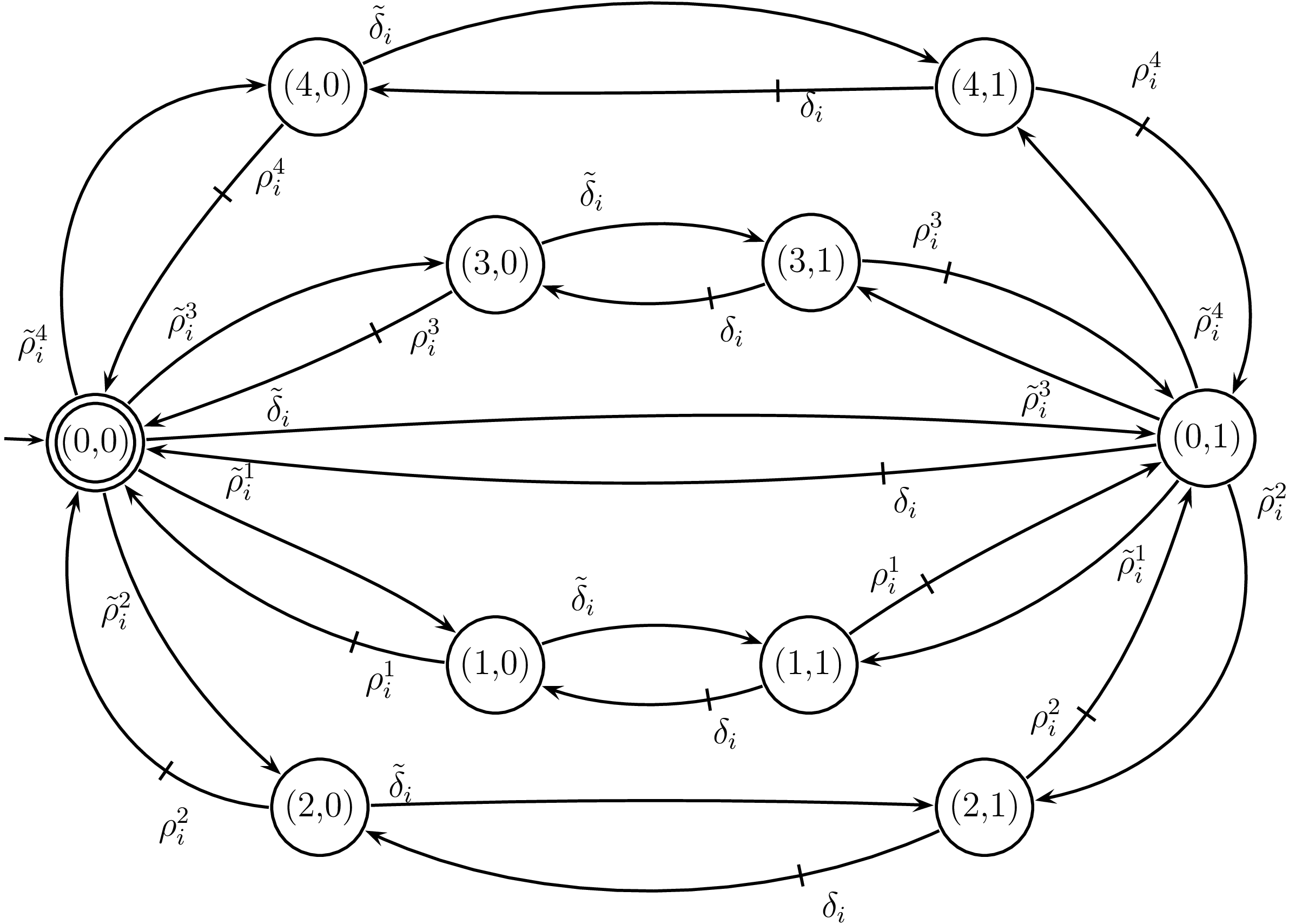}

\caption{Supervisors for adjacent spaces; the first index in each state represents the departures controller and the second index represents the arrivals controller. The initial state is the only marked state in $G_0$ and it represents the state where there are no aircraft waiting for permission to enter the airspace or take off.}
\label{fig:subplantOtherControllers}

\vspace{\abovecaptionskip}
\end{figure}

First, there is a permanent subplant that represents the supervisors for adjacent spaces, which is itself a synchronous product between subplants for departures and arrivals (Figure~\ref{fig:subplantOtherControllers}).

\begin{figure}[htb]
\centering
\captionsetup{captionskip=0pt,farskip=0pt,nearskip=0pt}

\vspace*{-\abovecaptionskip}

\includegraphics[width=0.65\textwidth,valign=T]{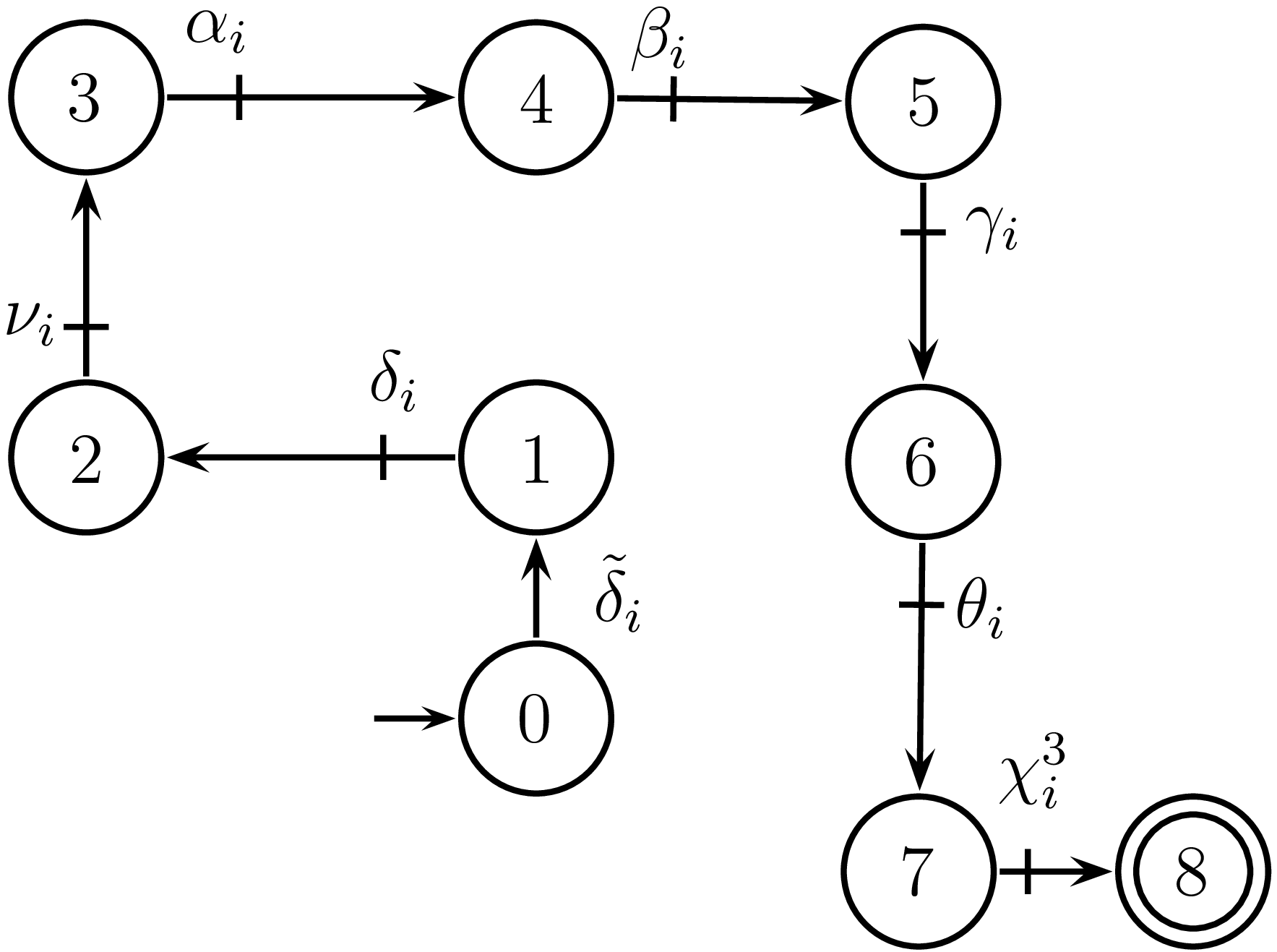}

\caption{A subplant representing an aircraft departing towards $\chi^3$. The initial state is before the aircraft has requested to depart and the marked state is when the the aircraft has successfully departed the airspace.}
\label{fig:aircraft}

\vspace{\abovecaptionskip}
\end{figure}

Second, there is a subplant for each aircraft that is airborne or ready to enter the airspace. These subplants will need to be added to or removed from the overall plant as aircraft arrive and depart from the airspace.  An example of an aircraft subplant is shown in Figure~\ref{fig:aircraft}; these subplants are denoted $G_i$ where $i$ is the aircraft's index. 

As in Chung et al.'s original paper, we assume that the language generated by the plant is equal to the prefix-closure of the plant's marked language: $L(G) = \overline{L_m(G)}$.\cite{Chung1992}

\subsection{The Specification}
Both the large state space of the problem and the plant's dynamic nature will make it difficult to write down the language generated by the plant, but we will also need to consider how we will describe the legal language, $K$. As in the original paper, we consider the legal language to be $L_m(G)$-closed, that is $K = \overline{K} \cap L_m(G)$.\cite{Chung1992}

At a high level, the supervisor must ensure that aircraft can take off from and land at this airfield with minimal restrictions and while guaranteeing flight safety. In plain English, the specifications captured by $K$ are:
 \begin{itemize}
 \item Two aircraft can't occupy the same section of airspace;
 \item An arriving aircraft can't be delayed for more than one controllable action;
 \item A departing aircraft can't be delayed for more than ten controllable actions; and
 \item Aircraft departing through the same gate must be separated by at least five actions.
 \end{itemize}
 
 \subsection{Summary}
It is clear from its formulation that the ATC problem meets all three of Chung et al.'s criteria for problems requiring LLP control.

First, there is an explosion in the size of the state space. If we consider the case of five aircraft airborne at once, the synchronous product of the subplants representing the adjacent supervisors and each aircraft has on the order of $6 \times 10^5$ states, i.e.\ even with a grossly simplified problem we have an impractical number of states.

Second, the plant is dynamic. Its composition changes every time an aircraft enters (or exits) the airspace. In the DES model this results in a subplant being added (or dropped) from the overall plant's synchronous product. This dynamic nature would require that an offline supervisor be recomputed for every new aircraft.

Third, the legal behaviour is difficult to capture as a regular language. This results from the plant's dynamic and modular natures: although event legality can be easily determined for any given point in time, determining this for all possible futures requires the supervisor to keep track of all possible legal languages.

\section{DES Operations for LLP Supervisors} \label{sec:DESoperations}
\newthought{All of the characteristics} of DES problems that make LLP supervisors necessary also make the analysis of the problems themselves somewhat difficult. We begin, therefore, by defining and illustrating some DES operations that will prove useful in our reasoning about the problem as well as the supervisor's own operations (all definitions from Chung et al.\cite{Chung1992}).

\begin{definition}[Active Set]
The active set\marginnote{The active set allows the supervisor to reason about what actions are possible in the plant from any given state.} in a language $L(G)$ after the trace $s$ is denoted $\Sigma_{L(G)}(s)$ and defined as $$\Sigma_{L(G)}(s) \coloneqq \{\sigma \in \Sigma | s\sigma \in L(G)\}.$$
\end{definition}
\begin{example}[Active Set] \label{ex:activeSetExample}
At any point in time where there are no aircraft airborne, the active set for the ATC problem is restricted to the alphabet of uncontrollable events. This is because the controllable events are only applicable when there is an aircraft in the airspace to be directed.
$$\Sigma_{L(G)}(\epsilon) = \Sigma_{uc}$$
\end{example}

\begin{definition}[Post-language]
The post-language\marginnote{The post-language operation provides the ``lookahead'' portion of LLP by allowing the future language of the plant to be determined on the basis of the history of the system. Since the possibility of some events is dependent on the state of the plant, this kind of operation cannot be done ahead of time and must be computed on-the-fly.} of a language $L$ after the trace $s$ is denoted $L/s$ and defined as $$L /s \coloneqq \{t \in \Sigma^* | st \in L\}.$$
\end{definition}

\begin{example}[Post-language] \label{ex:postLanguageExample}
If the aircraft from Figure~\ref{fig:aircraft} is the only aircraft airborne and $$s = \tilde{\delta}_1\delta_1\nu_1\alpha_1\beta_1\gamma_1$$ occurs, then the post-language contains strings such as
\begin{align*}
L(G)/s = \{&\theta_1,\theta_1\chi^3_1,\ \tilde{\rho}^{1}_2,\ \tilde{\rho}^{1}_2\theta_1,\ \tilde{\rho}^{1}_2\theta_1\rho^1_2,\  \tilde{\rho}^{1}_2\theta_1\rho^1_2\chi^3_1,\\
 &\tilde{\rho}^{1}_2\theta_1\rho^1_2\chi^3_1\alpha_2,\ \tilde{\rho}^{1}_2\theta_1\rho^1_2\chi^3_1\alpha_2\beta_2,\\
 &\tilde{\rho}^{1}_2\theta_1\rho^1_2\chi^3_1\alpha_2\beta_2\gamma_2,\ \tilde{\rho}^{1}_2\theta_1\rho^1_2\chi^3_1\alpha_2\beta_2\gamma_2\eta_2,\\
 &\tilde{\rho}^{2}_2,\ \tilde{\rho}^{2}_2\rho^2_2,\ \tilde{\rho}^{2}_2\rho^2_2\theta_1,\ \tilde{\rho}^{2}_2\rho^2_2\theta_1\gamma_2,\\
 &\tilde{\rho}^{2}_2\rho^2_2\theta_1\gamma_2\chi^3_1,\ \tilde{\rho}^{2}_2\rho^2_2\theta_1\gamma_2\chi^3_1\eta_2,\dots\}
\end{align*}
\end{example}

\begin{definition}[Truncation]
The truncation\marginnote{The truncation operation provides the ``limited'' portion of LLP by allowing the language of the plant to be trimmed to a tractable set of strings. As seen from Example~\ref{ex:postLanguageExample} it can be difficult to write strings out in full, let alone enumerate all possible strings. Instead, truncation restricts the language to those of its members whose length is no more than $N$.} of a language $L$ to the strings of length $N$ is denoted $L|_N$ and defined as $$L|_N \coloneqq \{t \in L\ |\ |t| \leq N\}$$ where $|t|$ is the length of the string $t$.
\end{definition}
\begin{example}[Truncation] \label{ex:truncationExample}
Extending Example~\ref{ex:postLanguageExample} the truncated post-language $L(G)/s|_3$ contains strings such as
\begin{align*}
\{&\theta_1,\ \theta_1\chi^3_1,\ \tilde{\rho}^{1}_2,\ \tilde{\rho}^{1}_2\theta_1,\ \tilde{\rho}^{1}_2\theta_1\rho^1_2,\\
&\tilde{\rho}^{2}_2,\ \tilde{\rho}^{2}_2\rho^2_2,\ \tilde{\rho}^{2}_2\rho^2_2\theta_1,\dots\}
\end{align*}
\end{example}

\section{Characterizing LLP Supervisors}
\begin{figure}[htb]
\centering
\captionsetup{captionskip=0pt,farskip=0pt,nearskip=0pt}

\vspace*{-\abovecaptionskip}

\includegraphics[width=\textwidth]{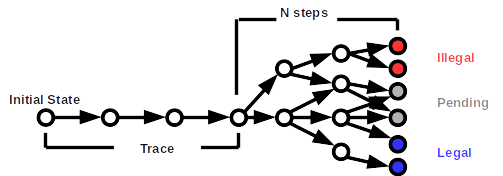}

\caption{An LLP supervisor projects the plant's behaviour for the next $N$ steps (adapted from Chung et al., 1992b)}
\label{fig:llpSupervisor}

\vspace{\abovecaptionskip}
\end{figure}

\newthought{With these useful definitions introduced}, we now turn our attention to formalizing the ideas behind LLP supervisors. An LLP supervisor (Figure~\ref{fig:llpSupervisor}) is a general supervisory framework and can be modelled in many different ways, nevertheless we formalize the operations in terms of automata, languages and strings as done by the original authors.\cite{Chung1992}

The LLP control scheme (Algorithm~\ref{alg:llpSupervisor}) aligns with traditional DES approaches: the biggest departure is in the third step where the controller must make assumptions about plant behaviour that is neither definitely legal nor definitely illegal. In practice the controller does this by adopting an attitude towards behaviour of uncertain legality: the controller can optimistically believe that all uncertain strings can be steered towards legality or the controller can conservatively restrict the plant to behaviour that is known to be legal.
 \begin{algorithm}
    \caption{LLP Supervisor}\label{alg:llpSupervisor}
    \begin{algorithmic}[1]
        \While{events occur in the plant}
        \State{Project the plant's behaviour over the next $N$ events $$L(G)/s|_N \text{ and } L_m(G)/s|_N)$$}
        \State{Determine legality of all strings $$\overline{K}/s|_N \text{ and } K/s|_N$$}
        \State{Label pending strings according to attitude $$f^N_a (s) = \begin{cases}
\text{cons. } & K/s|_{N-1}\\
\text{optm. } & K/s|_N \cup (\overline{K}/s|_N \setminus \overline{K}/s|_{N-1})\end{cases}$$}
        \State{Find the modified tree's supremal controllable sublanguage $$f^N(s) = [f^N_a(s)]^{\uparrow /s|_N}$$}
        \State{Output control action \label{eq:fuBlock}
$$\gamma^N(s) = \overline{f^N(s)|_1} \cup \overline{\Sigma_{uc} \cap \Sigma _{L(G)}(s)}$$}
        \EndWhile
    \end{algorithmic}
\end{algorithm} \marginnote[-10.25cm]{Algorithm~\ref{alg:llpSupervisor} is adapted from Chung et al., 1992b. The LLP control scheme is a five step process, with the supervisor performing the following calculations after each event: predict the possible future behaviour of the plant for the next $N$ steps; eliminate any traces the controller knows to be illegal; decide how to label the pending traces; calculate the pruned tree's supremal controllable sublanguage; and produce the control action.}

For the ATC problem this algorithm leads the LLP supervisor through the following determinations: whether aircraft might arrive and/or take off and where all aircraft in the airspace can move to; which of these behaviours respect the specifications and which do not; taking either a conservative or optimistic attitude towards behaviours that are not clearly legal or illegal; calculating the supremal controllable sublanguage of this pruned tree; and then outputting the control action for all aircraft under its control.

In order to be convinced that this formalization of an LLP supervisor leads to correct behaviour we must answer three questions:\cite{Chung1992}
\begin{enumerate}
\item What happens if the supervisor says that there is no policy that can keep the plant's behaviour legal?
\end{enumerate}
From a purely practical point of view, we are very interested in knowing how our supervisor will act when its calculations tell it that illegal behaviour cannot be prevented.
\begin{enumerate}
\setcounter{enumi}{1}
\item Is there a reason to prefer one attitude over the other? How differently do the optimistic and conservative attitudes behave?
\end{enumerate}
The two proposed attitudes seem to be suited to different scenarios. If this is true we would like to know how we can expect each attitude to behave and in which scenarios we should pick one attitude over the other.
\begin{enumerate}
\setcounter{enumi}{2}
\item Is it worth using an LLP supervisor instead of simply calculating an offline supervisor in the traditional manner?
\end{enumerate}
Given the weight of the literature showing that traditional supervisors are correct-by-construction and provably correct, we would like to know if we are trading away any of these guarantees when we decide to use an LLP supervisor. We will return to these three questions throughout the following presentation of Chung et al.'s technical results.

\subsection{Characteristic Error of the LLP Supervisor}
In answering the first question we get right to the heart of what distinguishes LLP supervisors. In contrast with a traditional offline supervisor, the characteristic and defining error of the LLP supervisor is the run-time error.
\begin{definition}[Run-time error]
If $s \in L(G,\gamma^N) \text{ and } f^N(s) = \emptyset,$ then we say that there is a \textbf{run-time error} happening in $L(G,\gamma^N)$ at trace $s$.
\end{definition}
A run time error exists when the LLP does not have any guaranteed safe action, in the sense that there is no trajectory from the current state of the system to a legal marked state that isn't at risk of being driven uncontrollably into illegal behaviour. The answer to the first question is therefore that if the supervisor says that no policy can prevent illegal behaviour in the plant then something has gone wrong and the supervisor should have prevented the plant from getting into this state in the first place.

A special case of the run-time error is when it occurs at the beginning of the plant's behaviour. This is called a starting error and it indicates that no supervisor can guarantee safe behaviour in the plant.

\begin{definition}[Starting error]
If a run-time error occurs for $s = \epsilon$, we call this a \textbf{starting error}.
\end{definition}

With this in mind, we should begin by asking whether the ATC problem has a starting error? If so, then it is not suitable for supervisory control and we must either reshape the problem or make use of other techniques.

\subsection{Validity of the LLP Supervisor}
The question remains whether a valid supervisor can be synthesized for the ATC problem with either the optimistic or conservative attitude. This is the question that we will focus on next, beginning with Chung et al.'s definition of supervisor validity.\cite{Chung1992}
\begin{definition}[Valid supervisor] \label{def:validSupervisor}
An LLP supervisor\marginnote{It is worth noting that the terminology of an LLP supervisor's \emph{validity} is potentially confusing given what it expresses. Clearly, a valid supervisor avoids blocking states as well as states which lead uncontrollably to illegal strings. Definition~\ref{def:validSupervisor} is stronger than this, however, and also requires that the supervisor be minimally restrictive.} with control policy $\gamma^N$ is called \textbf{valid} if $$L(G,\gamma^N) = \overline{K^\uparrow}.$$
\end{definition}
A valid supervisor for the ATC problem would therefore enforce flight safety, but only by restricting aircraft when necessary. The takeaway is that a valid LLP supervisor necessarily balances correctness with permissiveness.

\section{Attitudes for the LLP Supervisor} \label{sec:twoAttitudes}
\newthought{Chung et al.\ presented two possible attitudes} that LLP supervisors could take towards pending strings in their $N$-step lookahead windows: optimistic and conservative.\cite{Chung1992} We will consider the size of language produced by each attitude, the effect of larger windows on these language and whether these attitudes result in valid supervisors.

To begin, we state a result that will prove useful in our analysis. Put plainly, it states that if an attitude is at least as permissive for every possible string as another attitude, then the language allowed by the former attitude will be at least as big as the language allowed by the latter. This is stated more formally in Lemma~\ref{lem:technicalResult} and is proved by induction.\cite{Chung1992}
\begin{lemma} \label{lem:technicalResult}
If $\gamma_i(s) \supseteq \gamma_j(s)$ for all $s \in \Sigma^*$, then $L(G,\gamma_i) \supseteq L(G,\gamma_j)$.
\end{lemma}

\subsection{The Optimistic Attitude}
An LLP supervisor with the optimistic attitude\marginnote{This attitude will naturally allow every string in $\overline{K^\uparrow}$ to occur, since there will never be reason to believe that they will uncontrollably lead to an illegal state. It might also allow additional strings which do lead uncontrollably to illegal strings, however, since the information required to avoid these strings may lie outside the lookahead window.} marks all pending strings as legal, which means that it will presume that it can steer any trace to a string in $K$ until there is proof that it cannot.

\begin{example}[Uncontrollable illegality] \label{ex:uncontrollableIllegalityATC}
In the ATC problem, an optimistic supervisor could allow an aircraft in both the top-left and top-middle sections of airspace (Figure~\ref{fig:toptwoleft}).
\begin{figure}[htb]
\centering
\includegraphics[width=\textwidth]{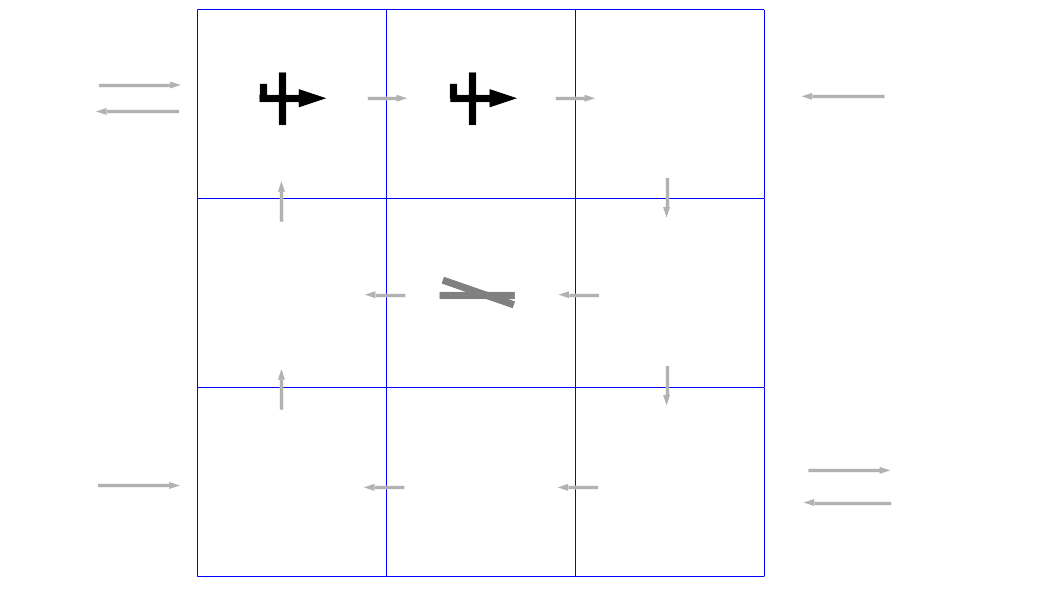}
\caption{Although strings that lead to this state are prefixes of legal, marked strings, the plant can lead uncontrollably to illegal behaviour if the event $\tilde{\rho}_i^{1}$ occurs.}
\label{fig:toptwoleft}
\end{figure}

This trace isn't illegal itself and is the prefix of a marked string, but it could uncontrollably lead to an illegal string: the event $\tilde{\rho}^{1}_i$ would require two controlled actions to avoid having two aircraft in the same section of airspace but that would contravene the specification that only one controlled event can happen before $\rho^1_i$.
\end{example}

\textbf{The language produced by the optimistic attitude} The optimistic attitude prioritizes maximal permissiveness over preventing illegal behaviour. The size of the language it produces, $L(G,\gamma^N_{optm})$, is provably bounded between the prefix-closure of the supremal controllable sublanguage of $K$ and the infimal controllable superlanguage of $K$.\cite{Chung1992}
\begin{theorem} \label{the:optmLangSize}
$\overline{K^\uparrow} \subseteq L(G,\gamma^N_{optm}) \subseteq K^\downarrow$
\end{theorem}

Intuitively, then, we hope that if we increase the size of the lookahead window then the LLP supervisor will have more of the information it needs to avoid illegal strings and that $L(G,\gamma^N_{optm}) = \overline{K^\uparrow}$ for some large enough $N$. Chung et al.\ proved this as Theorem~\ref{the:optmMonotone} which states that increasing the size of the lookahead window will never increase the size of the language allowed by an optimistic supervisor.\cite{Chung1992}
\begin{theorem} \label{the:optmMonotone}
$L(G,\gamma^N_{optm}) \supseteq L(G,\gamma^{N+1}_{optm})$
\end{theorem}

\textbf{Run-time errors with the optimistic attitude} An LLP supervisor with the optimistic attitude will never prohibit a string that is contained in $\overline{K^{\uparrow}}$, so we know that it will be minimally restrictive. We do have to verify, however, whether an illegal string might be allowed to occur. 
\begin{theorem} \label{the:optmValidity}
If a supervisor with the optimistic attitude is valid, then there will be no run-time errors in $L(G,\gamma^N_{optm})$.
\end{theorem}
\begin{proof}
Chung et al.'s proof\cite{Chung1992} of Theorem~\ref{the:optmValidity} is by contradiction and is broken down into two cases: when the current state's active set contains at least one uncontrollable action and, conversely, when the current state's active set is entirely controllable.

\begin{enumerate}
\item Consider when $\Sigma_u \cap \Sigma_{L(G)}(s) \neq \emptyset.$
\end{enumerate}
Because we assume the supervisor is valid and that a run-time error has occurred, we have that $f^N_{optm}(s) = K^{\uparrow}/s|_1 = \emptyset$. But, $\gamma^N_{optm}(s)$ must include the uncontrollable events that can occur in the plant, which makes it non-empty and a strict superset of $\overline{K^\uparrow}/s|_1$. This contradicts the definition of a valid supervisor, therefore $\gamma^N_{optm}$ is not valid.
\vskip0.25cm

\begin{enumerate}
\setcounter{enumi}{1}
\item Consider when $\Sigma_u \cap \Sigma_{L(G)}(s) = \emptyset.$
\end{enumerate}
Because we assume that a run-time error has occurred we have that $s \notin K$, otherwise the supervisor could permit $\epsilon$ to occur. If $\gamma^N_{optm}(s) = \emptyset$ and $s \notin K$ then our supervisor $\gamma^N_{optm}$ is blocking at $s$, which means that $s$ is not a prefix of any string in $K$. From this is follows that $L(G,\gamma^N_{optm}) \neq \overline{K^\uparrow}$ and therefore $\gamma^N_{optm}$ is not valid. 
\vskip0.25cm

For both cases, if a run-time error occurs then the supervisor was not valid, which is a contradiction. Therefore we conclude that if a supervisor with the optimistic attitude is valid, then there will be no run-time errors in $L(G,\gamma^N_{optm})$.
\end{proof}

This result is intuitive because by definition a valid supervisor is one that produces the prefix-closure of the supremal controllable sublanguage of the legal language. Unfortunately, the converse of Theorem~\ref{the:optmValidity} is not true. Chung et al.\ give a counterexample to show this, with the key being the potential for an unlimited number of uncontrollable events occurring.\cite{Chung1992}

\subsection{Conservative Attitude}
An LLP supervisor with the conservative attitude\marginnote{This attitude will never permit a string that is not part of the plant's supremal controllable sublanguage, but it may also prevent strings in $\overline{K^\uparrow}$ from occurring, since it may not be clear that their prefixes can be controlled all the way to a marked string.} marks all pending strings as illegal, which means that it presumes a trace leads uncontrollably to an illegal string unless there is proof that it does not.

\begin{example}[Overly restrictive]
In the ATC problem, this means that the supervisor could take more actions than necessary in order to ensure the legality of the produced strings.
\begin{figure}[htb]
\centering
\includegraphics[width=\textwidth]{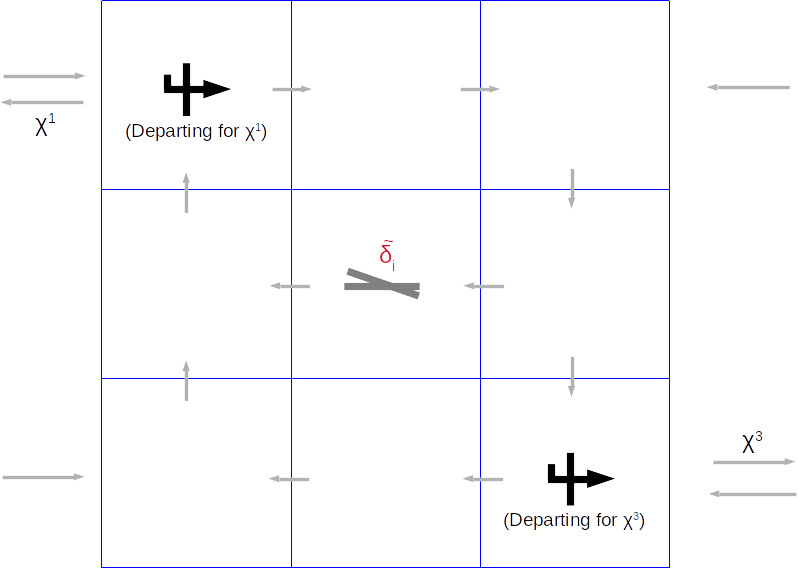}
\caption{Having received notification that an aircraft is ready to take off, the conservative attitude may lead a supervisor to prevent this until all the airborne aircraft have departed the airspace.}
\label{fig:delaydeparture}
\end{figure}

For example, consider a supervisor responsible for controlling the configuration shown in Figure~\ref{fig:delaydeparture} If this supervisor is notified that an aircraft is ready to depart, the event $\tilde{\delta}_i$ occurs, then the supervisor might unnecessarily delay that aircraft's departure until all airborne aircraft have departed the airspace.
\end{example}

\textbf{The language produced by the conservative attitude} The conservative attitude results in prioritizing the prevention of illegal behaviour over maximal permissiveness. The size of the language it produces, $L(G,\gamma^N_{cons})$, is provably bounded such that it is no larger than the prefix-closure of the supremal controllable sublanguage of $K$: as long as $\overline{K^\uparrow} \neq \emptyset$, a conservative supervisor composed with the plant is guaranteed to produce a subset of $\overline{K^\uparrow}$.\cite{Chung1992}
\begin{theorem} \label{the:consLangSize}
$K^\uparrow \neq \emptyset \iff L(G,\gamma^N_{cons}) \subseteq \overline{K^\uparrow}$
\end{theorem}

We would like to think, then, that by increasing the size of the lookahead window the LLP supervisor will be able to see that more strings in $\overline{K^\uparrow}$ can be permitted and we will have $L(G,\gamma^N_{cons}) = \overline{K^\uparrow}$ for some large enough $N$. Chung et al.\ formalize that increasing the lookahead window's size will never decrease the size of the language allowed by a conservative supervisor.\cite{Chung1992}
\begin{theorem} \label{the:consMonotone}
$L(G,\gamma^{N+1}_{cons}) \supseteq L(G,\gamma^N_{cons})$
\end{theorem}
\begin{proof}
Our proof follows Chung et al.'s from their addendum of technical results.\cite{Addendum1992} Due to Lemma~\ref{lem:technicalResult} the result in Theorem~\ref{the:consMonotone} will follow if we show $$f^{N}_{cons} \subseteq f^{N+1}_{cons}\quad \forall s \in \Sigma^*.$$
By the definition of the $f^N_u$ block (Algorithm~\ref{alg:llpSupervisor}, Line~\ref{eq:fuBlock}) we have $$f^N_{cons} \coloneqq (K/s|_{N-1})^{\uparrow/s|_N} = (K/s|_{N-1})^{\uparrow/s|_{N+1}}.$$
By the definition of truncation we have $$(K/s|_{N-1}) \subseteq (K/s|_N).$$
Since the post-language and supremal controllable sublanguage operations don't alter inclusion relationships
\begin{align*}
(K/s|_{N-1}) &\subseteq (K/s|_N)\\
&\implies (K/s|_{N-1})^{\uparrow/s|_{N+1}} \subseteq (K/s|N)^{\uparrow/s|_{N+1}}\\
&\implies f^{N}_{cons} \subseteq f^{N+1}_{cons}
\end{align*}
\end{proof}

\textbf{Run-time errors with the conservative attitude} From Theorem~\ref{the:consLangSize} we can see that as long as the plant's supremal controllable sublanguage is nonempty then the language produced by a supervisor with the conservative attitude will be nonempty as well.
\begin{theorem} \label{the:consValidity}
If there is no starting error in $L(G,\gamma^N_{cons})$, then there will be no run-time error in $L(G,\gamma^N_{cons})$.
\end{theorem}
\begin{proof}
Chung et al.'s proof\cite{Addendum1992} proceeds by induction on the length of the trace $s$.

The base case is $s = \epsilon$, where the length of $s$ is 0. There is no run-time error at $s$ because we have assumed that there is no starting error in $L(G,\gamma^N_{cons})$.

Our hypothesis is that no run-time error has occurred for $s$, a trace of length $i$, and we will show that this implies there is no run-time error for $s\sigma$, a trace of length $i+1$ where $\sigma \in \gamma^N_{cons}$.

Since there was no run-time error at $s$ we know that $f^N_{cons}(s) \neq \emptyset$; as a minimum $\sigma \in f^N_{cons}(s)$ since the conservative attitude allowed it to occur. With this in mind, as well as Theorem~\ref{the:consMonotone}'s result, if we consider the post-language $f^N_{cons}(s)/\sigma$ then we have $$\{\epsilon\} \subseteq f^N_{cons}(s)/\sigma = f^{N-1}_{cons}(s\sigma) \subseteq f^N_{cons}(s\sigma)$$ which means $f^N_{cons}(s\sigma) \neq \emptyset$ and no run-time error occurs at $s\sigma$. 
\end{proof}

Given its emphasis on enforcing legal behaviour, it makes sense that as long as the plant begins in a legal state, a supervisor with the conservative attitude will never allow the plant enter an illegal state.

\subsection{Comparing the two attitudes}
Having described the optimistic and conservative attitudes, we can ask ourselves how do these attitudes relate to each other, i.e.\ Chung et al.'s second question. We first consider the differences in the control actions they produce for one step ahead in the plant, formulated as Chung et al.'s first comparison result.\cite{Chung1992}
\begin{theorem}[Comparing One Step Ahead] \label{the:compareOne}
$\overline{f^M_{cons}(s)}|_1 \subseteq \overline{f^N_{optm}(s)}|_1\quad \forall\ s \in \Sigma^*,\quad \forall\ N,M \in \mathbb{N}$
\end{theorem}
Although this result is obvious when $N = M$, it is less clear that it should be true otherwise. With our previous results, however, the proofs are straightforward.
\begin{proof}
Consider three cases.
\begin{enumerate}
\item $N=M$. As stated above, this follows directly from the attitudes' respective definitions.
\item $N < M$. We have $$\overline{f^M_{cons}(s)}|_1 \subseteq \overline{f^M_{optm}(s)}|_1 \subseteq \overline{f^N_{optm}(s)}|_1$$ because increasing the lookahead window size does not increase the size of the language permitted by the optimistic attitude (Theorem~\ref{the:optmMonotone})
\item $N > M$. Similarly, we have $$\overline{f^M_{cons}(s)}|_1 \subseteq \overline{f^N_{cons}(s)}|_1 \subseteq \overline{f^N_{optm}(s)}|_1$$ because increasing the the lookahead window size does not decrease the size of the language permitted by the conservative attitude (Theorem~\ref{the:consMonotone})
\end{enumerate}
\end{proof}

This is a powerful relationship between the two attitudes and with Lemma~\ref{lem:technicalResult} it extends directly to comparing the languages produced by the two attitudes, formalized in Chung et al.'s second comparison result.\cite{Chung1992}
\begin{theorem}[Comparing Complete Languages] \label{the:compareLang}
$L(G,\gamma^M_{cons}) \subseteq L(G,\gamma^N_{optm})\quad \forall\ N,M \in \mathbb{N}$
\end{theorem}

The key takeaways from Theorem~\ref{the:compareLang} are that the choice of attitude has a definite effect on $L(G,\gamma^N)$ and that, unintuitively, this effect is independent of the size of lookahead window used. The power of this result is that we can be sure that the conservative attitude will never produce behaviour that is more permissive than the behaviour produced by the optimistic attitude, regardless of their respective lookahead window sizes.
\pagebreak
\section{Comparing Online and Offline Supervisors} \label{sec:CompareSupervisors}
\newthought{We will now address} Chung et al.'s third question: do we give up any guarantees or abilities when we forego an offline supervisor and we use an LLP supervisor instead?

The results covered in Section~\ref{sec:twoAttitudes} allow us to note that the sequences of languages $$\{L(G,\gamma^N_{cons})\}^\infty_{N=1} \text{ and }\{L(G,\gamma^N_{optm})\}^\infty_{N=1}$$ are both bounded and monotone, and we can therefore conclude that their limits, $L(G,\gamma^\infty)$, exist. As a final answer for the second question we would like to know if these limits are the same language and, to answer the third question, are these limits the same as the language produced by an offline controller? Formally, we are asking whether or not it is true that:
\begin{equation} \label{eq:limitOfLanguages}
\lim_{N \to \infty}L(G,\gamma^N) = \overline{K^\uparrow}
\end{equation}

Unfortunately, Equation~\ref{eq:limitOfLanguages} is not true in the general case, even under the assumption that $L_m(G)$ and $K$ are regular languages.\cite{Chung1992} All is not lost, however, since we can place conditions on the problem and on the value of $N$ that do guarantee that $L(G,\gamma^N) = \overline{K^\uparrow}$. This can be done whether or not $K$ is prefix-closed.

\subsection{Case I: $K$ is prefix-closed}
We start with the easier of the two cases, where the supervisor never has to worry about blocking, because the proofs are correspondingly simpler to follow.

To facilitate the analysis, Chung et al.\ introduce the measure $N_u(L)$ which is the length of the longest finite subtrace of uncontrollable events that occurs in language $L$.\cite{Chung1992}
\begin{equation}
N_u(L) \coloneqq  \begin{cases}
\max \{|s|\ |\ s \in \Sigma_{uc} \wedge \exists u,v \in \Sigma^*\\
 \left(usv \in L\right)\} \text{ if it exists}\\
\text{undefined otherwise}
\end{cases}
\end{equation}

\begin{figure}[htb]
\centering
\includegraphics[width=0.85\textwidth]{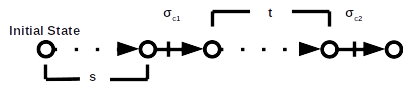}
\caption{Given that the trace $s$ has occurred in the plant, how big a lookahead window is needed to decide whether $\sigma_{c1}$ should be enabled or disabled?}
\label{fig:lookaheadWindow}
\end{figure}

\textbf{The optimistic attitude} Although we know that the absence of run-time errors does not imply that an optimistic supervisor is valid, we can make this claim if $K = \overline{K}$.\cite{Chung1992} Chung et al.\ proved that if $N$ can be made large enough to ensure no run-time errors in $L(G,\gamma^N_{optm})$ then it also ensures that $\gamma^N_{optm}$ is a valid supervisor.\cite{Chung1992}
\begin{theorem}[Conditions on $N$ -- Case I, Optimistic Attitude] \label{the:optmNI}
If $K = \overline{K}$ then
\begin{align*}
(N \geq N_u(L(G))+1) \ &\vee\ (N \geq N_u(K)+2)\\
& \implies L(G,\gamma^N_{optm}) = \overline{K^\uparrow}.
\end{align*}
\end{theorem}
Of note, for the first condition in Theorem~\ref{the:optmNI}'s antecedent we need only to look one event further than the longest finite uncontrollable subtrace that can be generated by the plant. As shown in Figure~\ref{fig:lookaheadWindow} if $t$ is the longest finite subtrace of uncontrollable events in $L(G)$, then an LLP supervisor with the optimistic attitude only needs to look $|t| + 1$ events ahead in order to allow $t$ to occur.

By contrast, in the second condition we must look two events further because the plant may be capable of generating longer uncontrollable subtraces than are included in the legal language. Consider in Figure~\ref{fig:lookaheadWindow} the case where $t$ is the longest finite subtrace of uncontrollable events in $K$. Since it is still possible that the plant could produce additional uncontrollable events after $t$, the optimistic LLP supervisor must look $|t| + 2$ events ahead in order to see that the plant can be controlled after $t$ occurs.

\textbf{The conservative attitude} Taking a different approach for the conservative attitude, we can show that if $L(G,\gamma^N_{cons})$ has no starting error then the conservative supervisor is valid.\cite{Chung1992} The key is the language
\begin{equation}
K^N_{pruned} = K \setminus (K/\Sigma_u^{N-1})\Sigma^*
\end{equation}
where if there is no starting error in $L(G,\gamma^N_{cons})$ and if $K \cap \Sigma^{(N-1)}_u = \emptyset$ then
\begin{equation}
L(G,\gamma^N_{cons}) = (K^N_{pruned})^\uparrow
\end{equation}

Chung et al.\ prove that if the plant begins in a legal state and $N$ can be made large enough to see the longest finite uncontrollable subtrace in $K$ (i.e.\ if $K^N_{pruned} = K$) then it also ensures that $\gamma^N_{cons}$ is a valid supervisor (Theorem~\ref{the:consNI}).\cite{Chung1992}
\begin{theorem}[Conditions on $N$ -- Case I, Conservative Attitude] \label{the:consNI}
If\marginnote{Because $K \subseteq L(G)$ we can use $N_u(L(G))$ to approximate $N_u(K)$. This transforms Theorem~\ref{the:consNI} into $$N \geq N_u(L(G))+2 \implies L(G,\gamma^N_{cons}) = \overline{K^\uparrow}$$ and is useful since it may be more practical to calculate $L(G)$ than $K$ for some problems.(Chung et al., 1992b).} $K = \overline{K}$ and there is no starting error in $L(G,\gamma^N_{cons})$ then
$$N \geq N_u(K)+2 \implies L(G,\gamma^N_{cons}) = \overline{K^\uparrow}.$$
\end{theorem}

It is notable that, unlike for the optimistic attitude, it is not sufficient that $N \geq N_u(L(G)) + 1$ to guarantee that $\gamma^N_{cons}$ is a valid supervisor. This is because if the longest uncontrollable subtrace in $L(G)$ appears in a string in $K$, then it is possible that $K \cap \Sigma_u^{N-1} \neq \emptyset$ and therefore $K^N_{pruned} \neq K$.

This is illustrated in Figure~\ref{fig:lookaheadWindow} by having a conservative LLP supervisor look $|t| + 1$ events ahead. Even if $t$ is the longest uncontrollable subtrace in $L(G)$ and even though the supervisor can see to the end of the subtrace in the plant's behaviour, the supervisor cannot see if any uncontrollable actions are possible after $t$ and will therefore conservatively prevent $\sigma_{c1}$ from occurring.

\textbf{Characterizing the ATC Problem} Theorems~\ref{the:optmNI} and~\ref{the:consNI} make it clear that, at a minimum, a problem only admits valid LLP supervisors if the plant cannot generate an infinitely long subtrace of uncontrollable actions. This is true of the ATC problem by construction, since the supervisors for adjacent spaces (Figure~\ref{fig:subplantOtherControllers}) have been restricted from putting our supervisor in an uncontrollably illegal state. The longest uncontrollable subtrace that can occur in the plant is of length two, e.g.\ $\tilde{\delta}_i\tilde{\rho}^{1}_j$, and after this subtrace occurs, the plant's active set is guaranteed to be restricted to controllable actions.

\subsection{Case II: $K$ is not prefix-closed}
The case where $K \neq \overline{K}$ is more difficult because our LLP supervisor must avoid blocking states as well as illegal states. Unfortunately, this is the relevant case for many real-world problems, including the ATC problem.

To analyze this case, Chung et al.\ introduce the idea of a ``frontier'' of legal states beyond which the plant can uncontrollably end up in an illegal state. This leads to the definition of two sets of traces: legal, marked, controllable traces; and uncontrollably crossing traces.\cite{Chung1992}
\begin{definition}[Legal, marked, controllable traces] The set of legal, marked, controllable traces is denoted $K_{mc}$ and is defined as
$$K_{mc} \coloneqq \{s \in K \cap L_c(G)\}$$ where the language $L_c(G)$ contains all the traces that can be produced by the plant after which the active set contains only controllable actions.$$L_c(G) \coloneqq \{s \in L(G)\ |\ s\sigma \notin L(G)\quad \forall\ \sigma \in \Sigma_{uc}\}$$
\end{definition} 
\begin{definition}[Uncontrollably crossing traces] The set of uncontrollably crossing traces is denoted $K_{\mathit{f\bar{c}}}$ and contains all the traces that lead from $\overline{K}$ to $L(G) \setminus \overline{K}$ on account of an uncontrollable event.
$$K_{\mathit{f\bar{c}}} \coloneqq \left(\left(L(G) \setminus \overline{K}\right)/\Sigma_{uc}\right) \cap \overline{K}$$
\end{definition}

\textbf{The optimistic attitude} The challenge for an optimistic supervisor is that it may allow traces to occur that lead to run-time errors. The key, then, is to ensure that the supervisor knows which strings in $K_{mc}$ are prefixes of strings in $K_{\mathit{f\bar{c}}}$, i.e.\ the supervisor must know which legal strings must be disallowed because they may uncontrollably lead the plant into illegal behaviour.

This is achieved by defining the measure $N_{\mathit{mcf\bar{c}}}$, which is the length of the longest subtrace $t$ that can occur after a legal, marked, controllable string $s$ such that $st$ is the only illegal string in $\overline{st}/s$.\cite{Chung1992}
\begin{equation}
N_{\mathit{mcf\bar{c}}} \coloneqq \begin{cases}
\max \{|t|\ |\ \exists s \in K_{mc} \cup \{\epsilon\} \left(st \in K_{\mathit{f\bar{c}}} \wedge\right.\\
\quad \left.\forall \epsilon < v < t \left(sv \notin K_{\mathit{f\bar{c}}} \cup K_{mc}\right)\right)\}\\
\quad\text{if it exists}\\
\text{undefined otherwise}
\end{cases}
\end{equation}

Chung et al.\ then prove that an LLP supervisor with the optimistic attitude is valid as long as $N$ is greater than $N_{\mathit{mcf\bar{c}}}$.\cite{Chung1992}

In Figure~\ref{fig:lookaheadWindow} consider the case where $s\sigma_{c1}$ is a legal string and $t$ is the longest subtrace that can occur such that $s\sigma_{c1}t$ is the only illegal string in the plant's post-language after $s\sigma_{c1}$. If $N < N_{\mathit{mcf\bar{c}}}+1$ then the LLP supervisor will not see the illegal state that occurs after $s\sigma_{c1}t$ and will optimistically allow the subtrace $t$ to occur. If $N \geq N_{\mathit{mcf\bar{c}}}+1$, on the other hand, the supervisor will have enough information to disable the event $\sigma_{c1}$ and prevent the subtrace $t$ from occurring.
\begin{theorem}[Conditions on $N$ -- Case II, Optimistic Attitude] \label{the:optmNII}
If $K^\uparrow \neq \emptyset$ then $$N \geq N_{\mathit{mcf\bar{c}}}+1 \implies L(G,\gamma^N_{optm}) = \overline{K^\uparrow}.$$
\end{theorem}

\textbf{The conservative attitude} In a similar fashion, the challenge for a conservative supervisor is that it may disallow traces that controllably lead to legal, marked states. The answer here is to ensure that the supervisor can see definitively whether or not every branch leads to a string in $K_{mc}$ or to an illegal string.

The measure to accomplish this is $N_{\mathit{mcmc}}$, which is defined as the length of the longest string $t$ that can occur after a legal, marked, controllable string $s$ such that $st$ is itself is the only legal, marked and controllable string in $\overline{st}/s$.\cite{Chung1992}
\begin{equation}
N_{\mathit{mcmc}} \coloneqq \begin{cases}
\max \{|t|\ |\ \exists s \in K_{mc} \cup \{\epsilon\} \left(st \in K_{mc} \wedge\right.\\
\quad\left.\forall \epsilon < v < t \left(sv \notin K_{mc}\right)\right)\}\\
\quad\text{if it exists}\\
\text{undefined otherwise}
\end{cases}
\end{equation}
This measure is used to formulate the guarantee in Theorem~\ref{the:consNII} namely that if $N$ is greater than $N_{\mathit{mcmc}}$ then a supervisor with the conservative attitude will have enough information to be a valid supervisor.\cite{Chung1992}

The reasoning that supports $N_{\mathit{mcmc}}$ as a measure is similar to the reasoning in support of $N_{\mathit{mcf\bar{c}}}$. In Figure~\ref{fig:lookaheadWindow} consider the case where $s\sigma_{c1}$ is a legal string and $t$ is the longest subtrace that can occur such that $s\sigma_{c1}t$ is the only legal string in the plant's post-language after $s\sigma_{c1}$. If $N < N_{\mathit{mcmc}}+1$ then the LLP supervisor will not see the legal state that occurs after $s\sigma_{c1}t$ and will conservatively prevent the subtrace $t$ from occurring. If $N \geq N_{\mathit{mcmc}}+1$, on the other hand, the supervisor will have enough information to allow the event $\sigma_{c1}$ and it will not needlessly restrict the legal string $s\sigma_{c1}t$ from occurring.
\begin{theorem}[Conditions on $N$ -- Case II, Conservative Attitude] \label{the:consNII}
If $\overline{K} = \overline{K_{mc}}$ and there is no starting error in $L(G,\gamma^N_{cons})$ then $$N \geq N_{\mathit{mcmc}}+1 \implies L(G,\gamma^N_{cons}) = \overline{K^\uparrow}.$$
\end{theorem}

\textbf{Characterizing the ATC problem} We recognize that the ATC problem's legal language, $K$, is not prefix-closed and that therefore Case II applies. To begin, we must check if the ATC problem meets the conditions for finding a value of $N$ where $L(G,\gamma^N) = \overline{K^\uparrow}$.
\begin{enumerate}
\item Does $K^\uparrow \neq \emptyset$?
\end{enumerate}
Unfortunately, no; the supremal controllable sublanguage of the ATC problem is empty. Without synthesizing the whole legal language $K$, we can consider the following example.
\begin{example}[Showing that $K^\uparrow = \emptyset$]
Our supervisor receives a notification that an aircraft will arrive from $\rho^1$ whenever possible; these notifications are uncontrolled events, since they are the result of other supervisors, and therefore can't be disabled. But if the trace $$s = \tilde{\rho}^{1}_1\rho^1_1\tilde{\rho}^{1}_2\alpha_1\rho^1_2\tilde{\rho}^{1}_3$$ occurs, then we will have aircraft in the top-left and top-middle sections of airspace and a third that must arrive to the top-left section after a maximum of one controllable action.
\begin{figure}[htb]
\centering
\includegraphics[width=\textwidth]{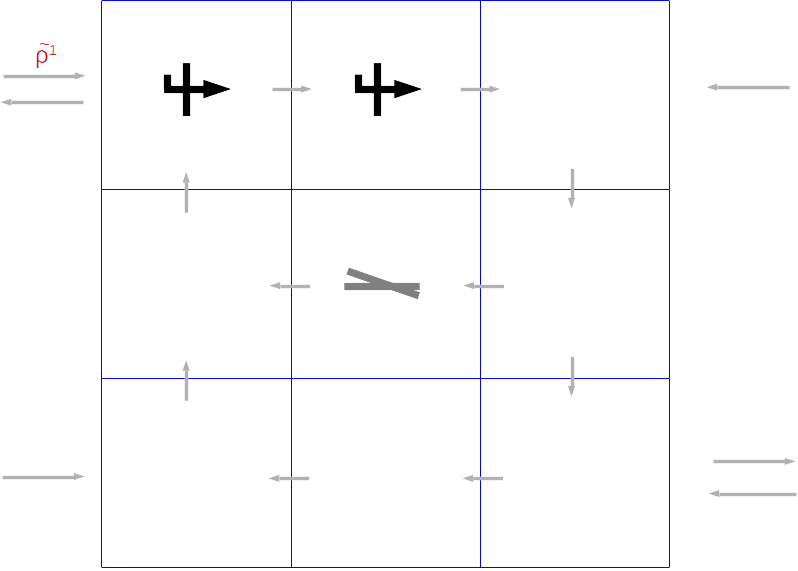}
\caption{The scenario in Example~\ref{ex:uncontrollableIllegalityATC} can be made to occur uncontrollably by the supervisor of the neighbouring airspace.}
\end{figure}

As seen in Example~\ref{ex:uncontrollableIllegalityATC} this cannot be done while respecting the specification that two aircraft cannot occupy the same section of airspace. The only controllable action that can occur after $s$ is $\beta_1$, but then both $\alpha_2$ and $\rho^1_3$ need to occur at the same time to prevent an illegal string.
\end{example}

With this, we conclude that the ATC problem as formulated does not admit a supervisor that can guarantee all the requirements are met, let alone such an LLP supervisor. If we make adjustments to the plant, however, we can ensure that adjacent supervisors do not force our supervisor into an illegal state. We formulate the following additional specifications and incorporate them into our plant:
\begin{itemize}
\item Departing aircraft must be separated by ten actions;\marginnote{The result is a new plant that has on the order of $10^7$ states but ensures that $K^\uparrow \neq \emptyset$.}
\item Aircraft arriving through the same gate must be separated by five actions; and
\item Only five aircraft can be in the airspace at a given time.
\end{itemize}

\subsection{Summary}
We have seen that, although in the general case $L(G,\gamma^\infty) \neq \overline{K^\uparrow}$, it is possible for some problems to choose a finite value for $N$ such that an LLP supervisor allows behaviour in the plant that is identical to that allowed by an offline supervisor.

To start, we must ensure some simple characteristics for the problem, namely that $K^\uparrow \neq \emptyset$ and that there is no starting error in $L(G,\gamma^N)$. The key is to calculate the appropriate measure, $N_u(K)$, $N_u(L(G))$, $N_{\mathit{mcf\bar{c}}}$ or $N_{\mathit{mcmc}}$, and to use this to set a lower bound on the size of the lookahead window. We have seen that these measure are undefined for some problems, however, which forces us to conclude that these problems do not admit a valid LLP supervisor with the given attitude.

\section{Synthesizing an LLP Supervisor} \label{sec:SynthesizeLLPSupervisor}
\newthought{Given that our application is safety-critical}, it is tempting to think that our LLP supervisor should implement the conservative attitude. Theorems~\ref{the:optmNII} and~\ref{the:consNII} make it clear, however, that both the optimistic attitude and conservative attitude can be used by a valid supervisor and that, in the limit, these result in the same language.

\subsection{An Optimistic LLP Supervisor}
In the first case, Theorem~\ref{the:optmNII} tells us that we having $N \geq N_{\mathit{mcf\bar{c}}} + 1$ will guarantee that an optimistic supervisor for the ATC problem will be valid. As has been discussed previously, the most dangerous situations that appear are when we allow aircraft to be arranged such that an arriving aircraft cannot be accommodated (see Example~\ref{ex:uncontrollableIllegalityATC}).
\begin{example}[$N_{\mathit{mcf\bar{c}}}$ for the ATC problem] \label{ex:calcNmcfc}
\begin{align*}
s = &\epsilon\\
t = &\tilde{\rho}^{3}_1\rho^{3}_1\kappa_1\lambda_1\mu_1\nu_1\tilde{\rho}^{3}_2\alpha_1\rho^3_2\kappa_2\lambda_2\mu_2\nu_2\\
&\tilde{\rho}^{3}_3\beta_1\rho^3_3\kappa_3\lambda_3\mu_3\alpha_2\tilde{\rho}^{3}_4\rho^3_4\gamma_1\kappa_4\lambda_4\\
&\tilde{\rho}^{4}_5\nu_3\rho^4_5
\end{align*}
We are required to take $s = \epsilon$ because $K \cap L_c(G) = \emptyset$ for the ATC problem, which implies that $K_{mc} = \emptyset$.

The string $t$ is the longest string that takes us from $s$ to an uncontrollably crossing trace; this is achieved by having four aircraft arrive from the gate with the furthest possible travel time and array themselves in a ``cross pattern'' (Figure~\ref{fig:exampleNmcfc})). Now any action except for $\eta_1$ runs into the problem that an uncontrollable notification regarding a fifth arriving aircraft can force us uncontrollably into illegal behaviour.
\begin{figure}[htb]
\centering
\includegraphics[width=\textwidth]{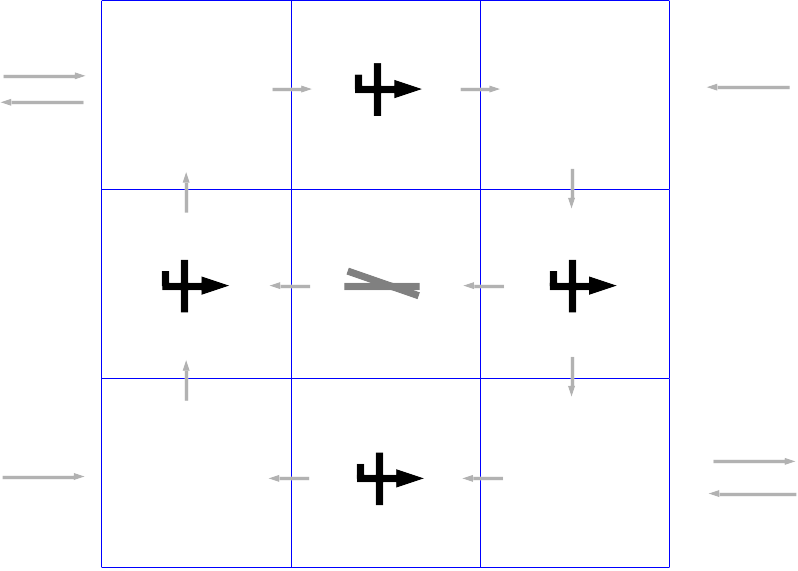}
\caption{The cross pattern.}
\label{fig:exampleNmcfc}
\end{figure}

Based on our string $t$, we have $N_{\mathit{mcf\bar{c}}} = |t| = 28$ and can therefore say that an optimistic LLP supervisor for the ATC problem should have a lookahead window of size $$N \geq 29.$$
\end{example}

\subsection{A Conservative LLP Supervisor}
Theorem~\ref{the:consNII} tells us that $N \geq N_{\mathit{mcmc}} + 1$ guarantees that a conservative LLP supervisor for the ATC problem will be valid. Unfortunately, as previously noted, $K_{mc} = \emptyset$ for the ATC problem which means that $N_{\mathit{mcmc}}$ is undefined. This reflects that at any point in the system, before we can reach a marked state we may uncontrollably have another aircraft turn up in the airspace. Borrowing terminology from other control theory domains, the ATC problem exhibits orbital stability and not asymptotic stability.

Interestingly, this means that we cannot find a valid conservative LLP supervisor with the current problem formulation. An obvious way to adjust the problem would be to mark all states representing safe behaviour throughout the plant in order to allow for the calculation of $N_{\mathit{mcmc}}$.
\begin{example}[$N_{\mathit{mcmc}}$ for modified ATC problem] \label{ex:calcNmcmc}
\begin{align*}
s = &\epsilon\\
t = &\tilde{\rho}^{3}_1\rho^{3}_1\kappa_1\lambda_1\mu_1\nu_1\tilde{\rho}^{3}_2\alpha_1\rho^3_2\kappa_2\lambda_2\mu_2\nu_2\\
&\tilde{\rho}^{3}_3\beta_1\rho^3_3\kappa_3\lambda_3\mu_3\alpha_2\tilde{\rho}^{3}_4\rho^3_4\gamma_1\kappa_4\nu_3\beta_2\lambda_4\mu_4\alpha_3\\
&\tilde{\rho}^{3}_5\nu_3\rho^3_5
\end{align*}
Although $K_{mc} \neq \emptyset$ for the modified ATC problem, $s = \epsilon$ still gives us the trace $t$ of maximum length.

The string $t$ is the longest string that takes us from $s$ to a string is $K_{mc}$; this is achieved by having four aircraft arrive and take as many actions as possible before a fifth aircraft arrives and prevents any uncontrollable actions from happening.

Based on $t$, we have $N_{\mathit{mcmc}} = |t| = 32$ and can therefore say that a conservative LLP supervisor for the modified ATC problem should have a lookahead window of size $$N \geq 33.$$
\end{example}

\subsection{Summary}
Of particular note in the last two sections is that, although the optimistic and conservative attitudes can both be used to produce $L(G,\gamma^N) = \overline{K^\uparrow}$, the particular value of $N$ required for each may be different. This results from the fact that the two attitudes need different kinds of information to produce valid supervisors; depending on the structure of the plant it may take more steps into the future for a given attitude to see what it needs to see.

\section{Extensions to Limited Lookahead Supervisors} \label{sec:LLPextensions}
\newthought{In addition to the material presented in their} seminal work,\cite{Addendum1992,Chung1992} the authors also published four notable extensions of LLP supervisors. These works introduced a technique to efficiently recompute the lookahead tree at every step, a more nuanced attitude towards pending strings, the ability to incorporate state information, and the ability to deal with partially observed plants.

\subsection{Efficient Computation of the Lookahead Tree}
The original authors presented a method for efficiently calculating the lookahead tree in a recursive manner, addressing a practical aspect of implementing the LLP control scheme.\cite{Chung1993}
\begin{algorithm}
\small
\caption{Recursive calculation for an LLP supervisor}
\label{alg:recursiveTreeComputation}
\begin{algorithmic}[1]
\Require{An $N$-step lookahead tree where $X^t(j)$ is the set of states at the $j$th level of the tree that was calculated after the $t$th event. $j = 1,...,N$ and $t=0,1,2,...$.}
\ForAll{States in $X^t(N)$} \label{alg:blockOneStart}
\If{Conservative attitude}
\State{The cost of this state is $\infty$.}
\ElsIf{Optimistic attitude}
\If{$x$ is an illegal state}
\State{The cost of this state is $\infty$.}
\Else{}
\State{The cost of this state is $0$.}
\EndIf
\EndIf
\EndFor \label{alg:blockOneEnd}
\For{Each layer, $j$, recursively back through the lookahead tree} \label{alg:blockTwoStart}
\ForAll{States in $X^t(j)$}
\If{Conservative attitude and this state's cost was $0$ in the last computed tree}
\State{The cost of this state is $0$.}
\ElsIf{Optimistic attitude and this state's cost was $\infty$ in the last computed tree}
\State{The cost of this state is $\infty$.}
\algstore{recursive1}
\end{algorithmic}
\end{algorithm}\marginnote[-8.75cm]{Algorithm~\ref{alg:recursiveTreeComputation} is adapted from Chung et al., 1993. To begin, the states in the $N$th layer of the tree have not been seen and are labelled according to the supervisor's attitude (Lines~\ref{alg:blockOneStart} to~\ref{alg:blockOneEnd}).

Finally, if the cost of the initial state is $0$, then the least restrictive control policy can be found using the ``cost-to-go'' values; if the cost of the initial state is $\infty$, then a run-time error has occurred (Lines~\ref{alg:blockThreeStart} to~\ref{alg:blockThreeEnd}).}

\begin{algorithm}
\begin{algorithmic}
\algrestore{recursive1}
\ElsIf{This state is marked or if this state is transient and there is a controlled action that leads to a state with cost $0$; and if no uncontrollable action leads to a state with cost $\infty$}
\State{The cost of this state is $0$}
\Else{}
\State{The cost of this state is $\infty$. }
\EndIf
\EndFor
\If{The cost of all states in this layer are $0$} \label{alg:terminationOne}
\State{Set the cost of all states in preceding layers to $0$ and terminate.}
\ElsIf{The cost of all states in this layer match their costs in the last computed tree} \label{alg:terminationTwo}
\State{Set the cost of all states in preceding layers to their costs in the last computed tree.}
\EndIf
\EndFor \label{alg:blockTwoEnd}
\If{The cost of the initial state is 0} \label{alg:blockThreeStart}
\State{\Return The least restrictive control policy.}
\Else{}
\State{\Return Run-time error}
\EndIf \label{alg:blockThreeEnd}
\end{algorithmic}
\end{algorithm}


This recursive calculation (Algorithm~\ref{alg:recursiveTreeComputation}) is based on converting the LLP control problem to a $0/\infty$ optimal control problem. In this problem, the cost of a policy $g(s)$ after trace $s$ has occurred in the plant is $0$ if $s$ is legal, if $g(s)$ permits all uncontrollable actions that can occur in the plant after the trace $s$, and if $g(s)$ is non-empty whenever $s$ is a prefix of a marked trace but not marked itself. All other policy/trace combinations are assigned a cost of $\infty$. This setup allows a dynamic programming approach to be used and makes it simple to apply when costs other than $0$ and $\infty$ are desired.

\subsection{Variable Lookahead Policies}
Chung et al.\ also extended their work by introducing variable lookahead policies (VLP) and the undecided cost, $U$. \cite{Chung1994} This cost is added to the optimal control framework that was introduced for recursively computing the lookahead tree, \cite{Chung1993} and is defined so that the following ordering exists amongst costs: $$0\quad <\quad U\quad <\quad \infty.$$

The power of the ``undecided'' cost is that pending strings can be assessed truthfully as opposed to arbitrarily labelling them based on supervisor attitude. This cost and its place in the ordering of costs reflect that an undecided string is not as desirable as a string with cost $0$, but it is certainly more desirable than a string with cost $\infty$. 

\begin{algorithm}[htb]
\small
    \caption{VLP supervisor labels pending string, $s$}\label{alg:vlpSupervisor}
    \begin{algorithmic}[1]
    \Require{A pending string $s$ that leads from the plant's current state to another state in the $N$-step lookahead tree.}
    \If{$|s| = N$} \label{alg:lengthNStart}
      \If{$s$ is not a prefix of a legal string}
	\State{Cost-to-go of $s$ is $\infty$}
      \Else
      \State{Cost-to-go of $s$ is $U$}
      \EndIf \label{alg:lengthNEnd}
    \ElsIf{$s$ leads to a marked, controllable state} \label{alg:insideStart}
      \State{Cost-to-go of $s$ is 0}
    \ElsIf{$s$ is not a prefix of a legal string}
      \State{Cost-to-go of $s$ is $\infty$}
    \Else
    \If{An uncontrollable event can occur}
    \State{Cost-to-go of $s$ is the maximum cost-to-go that occurs if an uncontrollable event follows $s$}
    \Else
     \State{Cost-to-go of $s$ is the minimum cost-to-go that occurs if a controllable event follows $s$}
     \EndIf
    \EndIf \label{alg:insideEnd}
    \end{algorithmic}
\end{algorithm}

Additional refinements presented included allowing the size of the lookahead window to be unbounded and again making intelligent reuse of previously computed trees unless the underlying plant has varied in the meantime. The authors presented experimental results from simulations and concluded that VLP supervisors expand significantly less of the lookahead tree than traditional LLP supervisors, that even a bounded VLP supervisor reduces the uncertainty in control policies, and that the unbounded VLP supervisor is the best supervisor on average.\marginnote[-11.75cm]{Algorithm~\ref{alg:vlpSupervisor} is adapted from Chung et al., 1994. Similar to the calculations in Algorithm~\ref{alg:recursiveTreeComputation}, a cost is assigned directly to $s$ if it leads to the edge of the lookahead tree (Lines~\ref{alg:lengthNStart} to~\ref{alg:lengthNEnd}). If the string leads to a state earlier on in the lookahead tree, though, then the costs-to-go of future states are used to determine the cost-to-go of the current state (Lines~\ref{alg:insideStart} to~\ref{alg:insideEnd}).}\cite{Chung1994}

\subsection{Using State Information}
Work up to date had assumed that little to no information was available regarding the underlying plant, although this allows easy generalization, it also ignores useful information.\cite{Hadj-Alouane1994} 
\begin{algorithm}
\small
    \caption{Online computation of control policy by VLP-S}\label{alg:vlpsSupervisor}
    \begin{algorithmic}[1]
    \Require{Initial state $x$ and underlying plant $G$}
    \State{Add all uncontrollable events that can occur in state $x$ to the control policy}
    \If{We don't know the cost-to-go for $x$} \label{alg:checkDomVOne}
    \State{Calculate the cost-to-go of $x$ using Algorithm~\ref{alg:vlpsCost}}
    \EndIf
    \If{The cost-to-go of $x$ is not $\infty$}
    \ForAll{Controllable actions $\sigma$ that can occur in state $x$}
    \If{We don't know the cost-to-go for $\delta(x,\sigma)$} \label{alg:checkDomVTwo}
    \State{Calculate the cost-to-go of $\delta(x,\sigma)$ using Algorithm~\ref{alg:vlpsCost}}
    \EndIf
    \If{The cost-to-go of $\delta(x,\sigma)$ is $0$}
    \State{Add $\sigma$ to the control policy}
    \EndIf
    \EndFor
    \EndIf
    \State{\Return the control policy}
    \end{algorithmic}
\end{algorithm}\marginnote[-6cm]{Algorithm~\ref{alg:vlpsSupervisor} is adapted from Hadj-Alouane et al., 1994. To calculate a state's cost-to-go, the lookahead window is first constructed by recursively expanding the plant $G$ from the current state $x$. This can be done with customizable stopping conditions to ensure that the procedure halts and that states are only expanded if necessary (Line~\ref{alg:recursiveExpand}). Once this is done, all of the illegal states, blocking states, and states that lead uncontrollably to these are assigned the cost-to-go $\infty$. Any remaining states are assigned the cost-to-go $0$, which allows the cost-to-go of the initial state to be returned.}

To address this, Hadj-Alouane et al.\ presented variable lookahead policies with state information (VLP-S), which allows reduced computation whenever states appear in multiple locations in the tree (Algorithm~\ref{alg:vlpsSupervisor}).\cite{Hadj-Alouane1994} The VLP-S supervisor constructs the control policy for the current state by allowing all uncontrollable events that can occur in the plant and then any controllable actions that lead to states with a cost-to-go of $0$.
\begin{algorithm}
\small
\caption{Calculating the cost-to-go of a state with VLP-S}
\label{alg:vlpsCost}
\algrenewcommand\algorithmicrepeat{\textbf{do}}
\algrenewcommand\algorithmicuntil{\textbf{while}}
\begin{algorithmic}[1]
\Require{State $x$ and underlying plant $G$}
\State{Recursively expand the subplant of $G$ from the state $x$} \label{alg:recursiveExpand}
\State{Assign all illegal states the cost-to-go $\infty$}
\Repeat
\State{Assign all states that lead uncontrollably to a state with cost-to-go $\infty$ the same cost-to-go}
\State{Assign all blocking states the cost-to-go $\infty$} \label{alg:blockingCost}
\Until{Any states were assigned a cost at Line~\ref{alg:blockingCost}}
\State{Assign all uncosted states the cost-to-go $0$}
\State{\Return the cost-to-go for state $x$}
\end{algorithmic}
\end{algorithm}\marginnote[-5.25cm]{Algorithm~\ref{alg:vlpsCost} is adapted from Hadj-Alouane et al., 1994.\\Algorithm~\ref{alg:vlpsSupervisor}, Lines~\ref{alg:checkDomVOne} and~\ref{alg:checkDomVTwo} are of particular note. The supervisor checks to see if the cost-to-go of a given state can be reused from a previous computation, which allows previously computed costs-to-go to be reused for future iterations of Algorithm~\ref{alg:vlpsSupervisor}.}

\subsection{Dealing with Partial Observation}
The last direct extension that we discuss is Hadj-Alouane et al.'s work to enable variable lookahead policies to consider plants under partial observation (VLP-PO).\cite{Hadj-Alouane1996} The authors present this as an extension of their previous work, VLP-S, to the specific case where the legal language is prefix-closed.

Since DES problems with partial observation\marginnote{Varying this ordering can lead to a different maximal sublanguage; the authors highlight that if this ordering is done with care then a desirable maximal sublanguage can be obtained depending on the domain-specific meaning of ``desirable.''} do not have supremal controllable and observable sublanguages, these policies generate instead a maximal controllable and observable sublanguage of the legal language on the basis of an argument ordering of the controllable events.\cite{Hadj-Alouane1996}

Only minimal changes are necessary\marginnote{The paper also presented a distributed version of VLP-PO and showed that it produced the same policies produced by the sequential VLP-PO. Both versions were demonstrated on a resource allocation problem, which allowed the authors to estimate that the distributed version would require a tenth of the run-time required by the sequential VLP-PO.}: first, the set of states that the plant could be in is computed based on the last observable event; then, the control action is computed in the manner of VLP-S. These computations are adapted to the partially observed nature of the plant and with respect to a global ordering provided for all events. Finally, the set of states that the plant could be in until the next observable event is computed, which will be an argument the next time that VLP-PO is called.
\pagebreak
\section{Additional Applications of Limited Lookahead} \label{sec:LLPapplications}
\newthought{Many other authors} have contributed to the literature on limited lookahead policies, building on Chung et al.'s broad conceptual base. We organize these into four categories (Table~\ref{tab:additionalApplications}) and cover them in the following sections.
\begin{table}[htb]
\caption{An overview of the literature investigating and applying LLP supervisors.}
\label{tab:additionalApplications}
\begin{tabular*}{\textwidth}{p{\textwidth}}
\toprule
\textbf{Different Underlying Plants} \\
Modifying an optimal controller to provide control for a partially observed system.\\
Decentralized and modular plants.\\
Probabilistic DES.\\
Hybrid systems.\\
Finite State Machines with Variables.\\
Partially observed Petri Nets with forbidden states.\\
Fault-tolerant supervisors.\\
\midrule
\textbf{Different Online Supervisors} \\
No explicit calculation of the supremal controlable sublanguage.\\
Near-optimal control of dynamic DES .\\
Enacting robust control.\\
Learning optimal LLP control using reinforcement learning.\\
\midrule
\textbf{Different Calculations} \\
Extending rather than truncating traces.\\
Estimating the lookahead tree's state-space size.\\
\midrule
\textbf{Applications} \\
Robots --- Navigation and task allocation.\\
 Software -- Fault detection and enforcing concurrency restraints.\\
\bottomrule
\end{tabular*}
\end{table}

\subsection{Different Underlying Plants}
Heymann and Lin noted that synthesizing supervisors for partially observed plants is an NP-hard problem and presented a process for modifying an optimal controller of the original, fully-observed system in an online fashion to provide control for a partially observed system. This process runs in $\mathcal{O}(n)$ time, where $n$ is the number of states contained in the the plant composed with the legal specification.\cite{Heymann1994} This was improved upon by Ushio, whose method always produces maximal controllable and observable sublanguages.\cite{Ushio1999}

For complex plants resulting from $n$ disjoint subsystems, Minhas and Wonham showed how modular calculations could be used to determine conditions on the size of the lookahead window.\cite{Minhas2003}  Online decentralized supervisors can be synthesized for these plants even if their total structure is unknown beforehand.\cite{Dai2014}

The underlying plant might also be probabilistic. Winacott and Rudie compared the performance of an LLP supervisory control approach named Recursive Utility Based Limited Lookahead (RUBLL) in this setting against the optimal solution derived using Markov Chain analysis. They concluded that RUBLL's performance was consistent with the optimal solution but was more readily applied to scenarios where exhaustive approaches are infeasible.\cite{Winacott2009}

For hybrid systems, such as a switching circuit, Dupuis and Fan compared an LLP supervisor's performance with that of an evolved finite state controller. They concluded that the LLP supervisor was better adapted to systems with slow dynamics and a vector field that flows towards the target, although their study only considered a 1-step lookahead policy for the LLP supervisor.\cite{Dupuis2010}

Finally, for partially observed Petri Nets Ru et al.\ synthesized supervisors that allow for control to be effected for any finite set of forbidden states.\cite{Ru2014}

\subsection{Different Online Supervisors}
Gu et al.\ propose a method for online synthesis of DES supervisors that divides the strings in a supervisor's lookahead tree into three types. Type I strings are those that include forbidden states and/or forbidden event strings; Type II strings include neither forbidden states nor forbidden event strings; and Type III strings include forbidden event strings when the past behaviour of the plant is considered. Obviously the strings of Types I and III are illegal, while those of Type II need to be tested for legality. The authors propose such a test based on the controllability of the related trace and claim that their method ``realizes optimal supervisory control without calculating the supremal controllable sublanguage.''\cite{Gu1996}

An optimal control framework is natural to consider if the supervisor must maximize the reward gained instead of guaranteeing legal plant behaviour. Grigorov and Rudie proposed a method to effect (near-)optimal control of dynamic DES\marginnote{A dynamic DES is one where the structure of the underlying plant is itself time-varying.}, considering aspects such as system reliability and normalizing for risk. A key takeway was that looking too far into the future and not far enough can both cause problems when planning.\cite{Grigorov2006}

Zhao et al.\ applied limited lookahead to deterministic finite state machines with variables and showed how they could enforce safety for a power grid in an online manner.\cite{Zhao2012}

Boroomand and Hashtrudi-Zad demonstrated that their Robust Limited Lookahead (RLL) supervisor could enact nonblocking robust control by taking a conservative attitude towards pending strings in the lookahead tree. They also showed that RLL resulted in maximally permissive policies if the lookahead window was of size $N_{nnf}$, the length of the longest neighbouring nonblocking frontier trace.\cite{Boroomand2013a}

In order to allow for supervisors to adapt as they enact control, Umemoto and Yamasaki showed that a supervisor could learn an optimal LLP by using reinforcement learning. Specifically, for a non-stationary environment the LLP supervisor is able to learn the state transition probabilities, expected rewards and expected costs of enacting control. The authors conducted simulations that showed how different learning rates trade convergence rate for asymptotic performance.\cite{Umemoto2015}

For situations where uncontrolled faults can occur in the plant, Dai et al.\ considered whether limited lookahead could be used to achieve fault-tolerance in supervisors. Their proposed method contained three parts: a learned nominal supervisor; a learned fault detector; and a post-fault supervisor. Necessary and sufficient conditions are presented for the existence of a satisfactory post-fault supervisor.\cite{Dai2016}

\subsection{Different Calculations}
Kumar et al.\ approached the idea of LLP from a notion of extending rather than truncating, as Chung et al.\ and most other surveyed works had done. The result of this approach is an extension based Limited Lookahead (ELL) supervisor, which estimates the plant behaviour by adding all finite-length event sequences to the N-step projection of plant behaviour. The key benefit of this approach is that it obviates the need for the supervisor to take an attitude by ensuring that no pending strings occur while requiring the same order of computation as the traditional LLP supervisor.\cite{Kumar1998}

Addressing a topic of practical concern, Winacott et al.\ presented and analysed a method for estimating the state-space size for an LLP supervisor's lookahead tree. This method is based on the underlying plant's adjacency matrix and depends on estimating a parameter $\tau$ that captures the state space's size.\cite{Winacott2013}

\subsection{Applications}
Besides theoretical developments, there have also been a number of works applying LLP theory to specific problem domains.

For robotics, Kobayashi and Ushio applied LLP to a Petri net model of mobile robots navigating through a building,\cite{Kobayashi2000} while Tsalatsanis et al.\ used LLP to dynamically allocate tasks to a team of robots.\cite{Tsalatsanis2012}

For software applications, Zhao et al.\ used limited lookahead DES to actively monitor software in order to predict faults and avoid them before they occurred.\cite{Zhao2010} Auer et al.\ considered the problem of multithreaded applications and applied limited lookahead to guarantee that concurrency restraints were enforced.\cite{Auer2014}.

\section{Conclusion}
\newthought{In this tutorial} we introduced the basic terminology required when discussing LLP supervisors. Although the underlying ideas for LLP supervisors are intuitive, we presented Chung et al.'s formal proofs to rigorously confirm their correctness. We demonstrated that both the optimistic and conservative attitudes can be used to produce valid supervisors as well as the conditions for which this is the case.

The ATC problem illustrated Chung et al.'s ideas throughout the tutorial, including: characterizing the underlying problem, recognizing when the problem did not allow for LLP control and highlighting that the required window size may vary based on the supervisor's attitude. The ATC problem also illuminated some possible routes for developing the theory of LLP supervision, including the idea that some plants may exhibit a kind of orbital stability. In such cases it would be useful to bound the window size $N$ which guarantees that the supervisor can prevent illegal behaviour and ensure that each of its subplants reaches their respective marked states in finite time without requiring that the plant as a whole reaches its marked state.

Finally, we reviewed extensions to Chung et al.'s original paper, many by the original authors, as well as applications of this theory to a wide variety of specific problems and problem classes. We believe that this tutorial will enable other DES practitioners to make use of LLP supervisors when appropriate in their own work.

\section*{Acknowledgments}
The authors acknowledge that Queen's University is situated on traditional Anishinaabe and Haudenosaunee Territory.
\vspace{0.5cm}

All DFA figures were produced using the Integrated\marginnote{IDES is available under the AGPL-3.0 open source license at \url{https://github.com/krudie/IDES}.} Discrete Event Systems (IDES) software.\cite{Rudie2006}
\vspace{0.5cm}

This research was supported by the Natural Sciences and Engineering Research Council of Canada as well as the Faculty of Engineering and Applied Science at  Queen's University.

\small
\bibliographystyle{plainnat}
\bibliography{./LookaheadTutorial.bib}

\end{document}